\documentclass[aps,prd,twocolumn,showpacs,nofootinbib]{revtex4-1}

\usepackage{graphicx}
\usepackage{amsmath}
\usepackage{amssymb}
\usepackage{bm}
\usepackage{bbm}
\usepackage{color}

\begin{document}

\title{Next-to-leading order QCD corrections for the charmonium production via the channel $e^+ e^- \to H(|c\bar{c}\rangle) + \gamma$ round the $Z^0$ peak}

\author{Gu Chen}
\author{Xing-Gang Wu}
\email{email:wuxg@cqu.edu.cn}
\author{Zhan Sun}
\author{Xu-Chang Zheng}
\author{Jian-Ming Shen}

\address{Department of Physics, Chongqing University, Chongqing 401331, P.R. China}

\date{\today}

\begin{abstract}

In our previous work~\cite{zcharmlo}, it has been found that sizable charmonium events can be produced via the channel $e^+e^- \to \gamma^*/Z^0 \to H(|c\bar{c}\rangle) +\gamma$ at the suggested super $Z$ factory, where $H(|c\bar{c}\rangle)$ represents the dominant color-singlet $S$-wave and $P$-wave charmonium states $J/\psi$, $\eta_c$, $h_c$ and $\chi_{cJ}$ ($J=0, 1, 2$), respectively. As an important step forward, in the present paper, we present a next-to-leading order (NLO) QCD analysis within the framework of nonrelativistic QCD. In different to the case of $B$ factory in which the single charmonium production is dominated by the channel via a virtual photon, at the super $Z$ factory, its cross-section is dominated by the channel via a $Z^0$ boson. Estimations up to NLO level are done under the condition of both the $B$ factory and the super $Z$ factory. We observe that the NLO distributions have the same shapes as those of LO distributions, but their differences are sizable. This indicates that a NLO calculation is necessary and important to achieve a more accurate estimation. Due to $Z^0$ boson resonance effect, at the super $Z$ factory with a high luminosity up to $10^{36}{\rm cm}^{-2}{\rm s}^{-1}$, when summing all the color-singlet states' contribution together, one may observe about $8.0\times10^{4}$ charmonium events via the channel $e^+e^-\to Z^0 \to H(|c\bar{c}\rangle)+\gamma$ in one operation year. Then, such super $Z$ factory could provide another useful platform to study the charmonium properties, even for the higher charmonium states.

\end{abstract}


\maketitle

\section{Introduction}

Heavy quarkonium is a multiscale system which provides an ideal platform for probing quantum chromodynamics (QCD) theory at all energy regions. The charmonium production at the $e^+ e^-$ collider via exclusive processes can be helpful for the purpose. At the $B$ factories as Belle and \textsl{BABAR}, which are running with the center-of-mass collision energy $\sqrt{s}=10.6$ GeV, the single charmonium production is dominated by the channel via a virtual photon, i.e. $e^+ e^- \to \gamma^* \to H(|c\bar{c}\rangle)+\gamma$, where $H(|c\bar{c}\rangle)$ stands for the $S$-wave or $P$-wave color-singlet charmonium, respectively. Because only one charmonium in the final state, one can extract more subtle properties of the charmonium. In the literature, the cross section of this channel has been studied up to next-to-leading order (NLO), c.f. Refs.~\cite{zcharmlo,higher-charm2,prdDL,higher-charm}. It has been found that the cross sections for the single charmonium production are larger than those of double charmonium production by about several times up to an order of magnitude. Since the double charmonium channel such as $e^+ e^- \to \gamma^* \to J/\psi+\eta_c$ has already been measured at the $B$ factories~\cite{Teva,Tevb,BABAR}, it has been optimistically estimated that if the background from the channel $e^+e^-\to X+\gamma$ is under well control in the recoil mass region near the $H(|c\bar{c}\rangle)$ resonance, the process $e^+e^-\to\gamma^*\to H(|c\bar{c}\rangle)+\gamma$ can also be detected by analyzing the photon energy spectrum in $e^+e^-\to X+\gamma$. However, till now, there is no experimental observation about the associated production from those two $B$ factories. It is therefore helpful to find another experimental platform to check all theoretical estimations. And a super $Z$ factory running at an energy around the $Z^0$-boson mass with a high luminosity ${\cal L}\simeq 10^{34-36}{\rm cm}^{-2}{\rm s}^{-1}$~\cite{zfactory} could be a good candidate for such purpose.

Considering the high luminosity and a clean environment of the super $Z$ factory, more and more rare decays and productions can be observed and measured. At the super $Z$ factory, the single charmonium production is dominated by the channel via $Z^0$ boson, i.e. $e^+ e^- \to Z^0 \to H(|c\bar{c}\rangle)+\gamma$, due to the $Z^0$-boson resonance effect~\cite{zcharmlo,higher-charm2}. More explicitly, it shows that for high luminosity ${\cal L}\simeq 10^{36}{\rm cm}^{-2}{\rm s}^{-1}$, we shall have around $5.0\times10^4$ $J/\psi$, $7.5\times10^3$ $\eta_c$, $6.2\times10^3$ $h_{c}$, $3.3\times10^3$ $\chi_{cJ}$ events for the channel of $e^+e^- \to Z^0 \to H(|c\bar{c}\rangle) +\gamma$ by one operation year \cite{zcharmlo}. Thus, such a super $Z$ factory can be an useful platform for studying heavy quarkonium properties and for testing QCD theories. As a sound estimation, it is interesting to show how the NLO corrections affect the previous leading-order estimations, as is the purpose of the present paper. In addition to previous NLO corrections to the channel $e^+ e^- \to \gamma^* \to H(|c\bar{c}\rangle)+\gamma$ done in Refs.~\cite{prdDL,higher-charm} and at the condition of $B$ factory, we shall also deal with the channel $e^+ e^- \to Z^0 \to H(|c\bar{c}\rangle)+\gamma$ up to NLO level. We shall present a sound estimation at both the $B$ factory and the super $Z$ factory.

The remaining parts of the paper are organized as follows. In Sec.II, we present a NLO calculation on the single charmonium production via the channel $e^+ e^- \to \gamma^*/Z^0 \to H(|c\bar{c}\rangle)+\gamma$ within the framework of nonrelativistic QCD (NRQCD)~\cite{nrqcd}. Numerical results and discussions are presented in Sec.III. The final section is reserved for a summary. Analytical expressions for basic one-loop integrations are put in the Appendix.

\section{Calculation technology}

In this section, we describe our calculation technology for dealing with the single charmonium production via the channel $e^+ e^- \to \gamma^*/Z^0 \to H(|c\bar{c}\rangle)+\gamma$ up to NLO level. Because the color-octet components provide negligible contributions, we shall concentrate our attention on the charmonium production via the color-singlet mechanism~\cite{zzx}. Since the LO part has been detailed analyzed in Ref.\cite{zcharmlo}, we shall mainly provide the technology on how to deal with the NLO part and shall list some of the LO results only for self-consistence.

\begin{figure}[htb]
\includegraphics[width=0.45\textwidth]{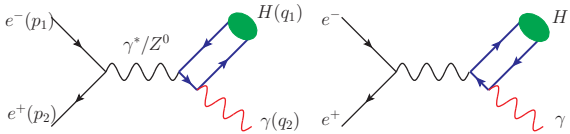}
\caption{ Tree-level diagrams for $e^+(p_{2})+e^-(p_{1})\to \gamma^{*}/Z^{0} \to H(|c\bar{c}\rangle)(q_{1})+\gamma(q_{2})$, where $H(|c\bar{c}\rangle)$ stands for color-singlet $S$-wave or $P$-wave charmonium states: $|[c\bar{c}]_{\bf 1}(^1S_0)\rangle$, $|[c\bar{c}]_{\bf 1}(^3S_1)\rangle$, $|[c\bar{c}]_{\bf 1}(^1P_1)\rangle$ and $|[c\bar{c}]_{\bf 1}(^3P_J)\rangle$ with $J=0,1,2$, respectively. }  \label{fig1}
\end{figure}

\begin{figure*}
\includegraphics[width=0.9\textwidth]{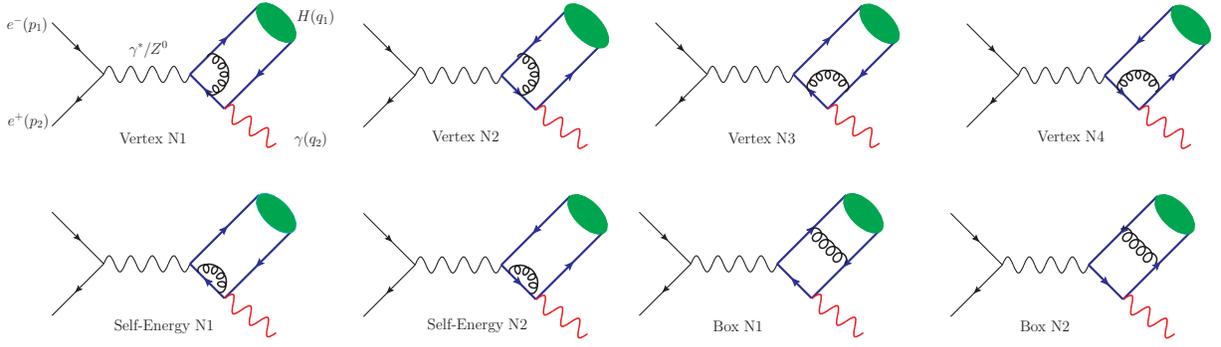}
\caption{One-loop diagrams for $e^+(p_{2})+e^-(p_{1})\to \gamma^{*}/Z^{0} \to H(|c\bar{c}\rangle)(q_{1})+\gamma(q_{2})$, where $H(|c\bar{c}\rangle)$ stands for color-singlet $S$-wave or $P$-wave charmonium states: $|[c\bar{c}]_{\bf 1}(^1S_0)\rangle$, $|[c\bar{c}]_{\bf 1}(^3S_1)\rangle$, $|[c\bar{c}]_{\bf 1}(^1P_1)\rangle$ and $|[c\bar{c}]_{\bf 1}(^3P_J)\rangle$ with $J=0,1,2$, respectively. }
\label{fig2}
\end{figure*}

Up to NLO accuracy, there are two tree-level Feynman diagrams and eight one-loop diagrams, which are shown in Figs.~\ref{fig1} and \ref{fig2}. It is noted that for the present color-singlet charmonium production, at the NLO level, there are only virtual corrections and the real corrections have no contribution to the amplitudes due to their color factors vanish, $\textrm{Tr}\left[\frac{T^a}{\sqrt{3}}\right] = 0$ with $T^a$ being the color factor for the color-octet charmonium. Then, the differential cross section for the process $e^+(p_{2})+e^-(p_{1})\to \gamma^{*}/Z^{0} \to H(|c\bar{c}\rangle)(q_{1})+\gamma(q_{2})$ at the NLO level can be schematically written as,
\begin{eqnarray}
\textrm{d}\sigma = \textrm{d}\sigma_{Born} + \textrm{d}\sigma_{Virt} +{\cal O}(\alpha_s^2) \;,
\end{eqnarray}
where,
\begin{equation}
\textrm{d}\sigma_{Born} = \frac{1}{2s} \overline{\sum} |{\cal M}_{Born}|^{2} \textrm{d}\Phi_2
\end{equation}
and
\begin{equation}
\textrm{d}\sigma_{Virt} = \frac{1}{2s} \overline{\sum} 2\textrm{Re}({\cal M}_{Born}{\cal M}^*_{Virt}) \textrm{d}\Phi_2 \;,
\end{equation}
where $s=(p_{1}+p_{2})^2$. The symbol $\overline{\sum}$ means averaging over initial states and summing over the final ones. ${\cal M}_{Born}$ and ${\cal M}_{Virt}$ are amplitudes for the Born level and the virtual corrections, respectively. $\textrm{d}\Phi_n$ is the $n$-body phase space which is formulated as,
\begin{equation}
\textrm{d}{\Phi _n} = {(2\pi )^4} {\delta ^4}\left({p_1} + {p_2} - \sum\limits_{f=1}^n {{q_f}} \right) \prod \limits_{f = 1}^n {\frac{\textrm{d}\vec{q}_{f}} {(2\pi)^3 2q_f^0}} \;.
\end{equation}
Details to deal with the phase space, e.g., to get the total cross sections and to get the differential cross sections for angular distribution and transverse momentum distribution, can be found in Ref.~\cite{zcharmlo}.

Different to previous treatment of using improved trace technology~\cite{itt0,zbc0,tbc,zbc1,zbc2,eebc,wbc} to deal with the hard scattering amplitude for the channel $e^{+}+e^{-}\to \gamma^{*}/Z^{0} \to H(|c\bar{c}\rangle)+\gamma$~\cite{zcharmlo}, because of its complexity at the NLO level~\footnote{At present, we are trying to to extend the improved trace technology up to NLO level, which is in progress.}, we shall adopt the conventional squared trace technology to do the calculation. For the purpose, we adopt the {\tt FeynArts} package~\cite{feynarts,feynarts1} to directly generate Feynman diagrams and Feynman amplitudes. And the followed basic color algebra and Dirac matrix simplifications are done by using {\tt FeynCalc} package~\cite{FeynCalc}.

There are ultraviolet (UV), infrared (IR), and Coulomb singularities in ${\cal M}_{Virt}$. The UV divergences will generally appear in self-energy and triangle diagrams, while the box diagrams are free of UV divergence. The triangle and box diagrams are in general IR divergent. More specifically, for the processes $e^+e^- \to \gamma^{*}/Z^{0} \to H(|c\bar{c}\rangle)+\gamma$, the self-energy diagrams labeled by Self-Energy N1 and Self-Energy N2 in Fig.~\ref{fig2} are UV divergent but IR finite; the Box N1 and Box N2 are UV safe but IR divergent and also have Coulomb singularities which should be absorbed into the non-perturbative color-singlet matrix elements; the triangle diagrams are both UV and IR divergent. After applying the dimensional renormalization~\cite{dr} with the space-time dimension $D=4-2\epsilon$, these infinities can be safely removed and one could obtain finite results.

For the process, $e^+e^- \to \gamma^{*}/Z^{0} \to H(|c\bar{c}\rangle) +\gamma$, there are two renormalization constants, $Z_m$ and $Z_2$, which correspond to charm quark mass $m_c$ and charm field $\psi_c$, respectively. We adopt the on-mass-shell (OS) scheme to set the renormalization constants $Z_m$ and $Z_2$, which satisfy
\begin{widetext}
\begin{eqnarray}
\delta Z_2^{OS} &=& -C_F \frac{\alpha_s}{4\pi} \left[ \frac{1}{\epsilon_{\textrm{UV}}} + \frac{2}{\epsilon_{\textrm{IR}}}-3\gamma_{E}+3\ln\frac{4\pi\mu^2_R}{m^2_c}+4 \right]+{\cal O}(\alpha_s^2),\\
\delta Z_m^{OS} &=& -3C_F \frac{\alpha_s}{4\pi} \left[ \frac{1}{\epsilon_{\textrm{UV}}} - \gamma_{E} + \ln\frac{4\pi\mu^2_R}{m^2_c}+\frac{4}{3} \right]+{\cal O}(\alpha_s^2) ,
\end{eqnarray}
\end{widetext}
where $\mu_R$ stands for the renormalization scale and $\gamma_{E}=0.577$ is the Euler constant. Then, one can obtain all counter terms analytically and the UV divergences can be canceled by adding all those counter terms together.

In evaluating the amplitude ${\cal M}_{Virt}$, we have to deal with the divergent loop integrals. We take the diagram Box N1 of Fig.~\ref{fig2} as an explanation of how to deal with those integrals. Its amplitude can be formulated as
\begin{equation} \label{9}
{\cal M}^{\textrm{B1}}_{Virt} ={\cal C} \times {{{\bar{v}}_{s'}}(p_2) L^{\mu} u_{s}(p_1)} D_{\mu \nu} I^{\nu \sigma} \varepsilon_{\sigma}(q_2) \;.
\end{equation}
The superscript $\textrm{B1}$ denotes the diagram Box N1 shown in Fig.~\ref{fig2}, similarly, the self-energy and triangle diagrams can be noted as $\textrm{S}1$ and $\textrm{S}2$, and $\textrm{V}1 \cdots \textrm{V}4$, respectively. ${\cal C}$ is the color factor. The subscripts $s$ and $s'$ represent the spin projections of the initial particles. $L^{\mu}$ is a Dirac $\gamma$-string and the $D_{\mu\nu}$ is the propagator for virtual photon or $Z^0$ boson. For the process via a virtual photon, we have
\begin{eqnarray}
{L^\mu } =  - ie{\gamma ^\mu }, \; D_{\mu\nu}=\frac{-ig_{\mu\nu}}{k_1^2},
\end{eqnarray}
and for the process via a $Z^0$ boson propagator, we have
\begin{eqnarray}
{L^\mu } &=& \frac{{i g_w}} {{4\cos {\theta _w}}}{\gamma ^\mu }(1 - 4{\sin ^2}{\theta_w} - {\gamma ^5}) \;, \\
D_{\mu\nu} &=& \frac{i}{k_1^2-m^2_Z +im_Z\Gamma_z}\left(-g_{\mu\nu}+{k_{1\mu} k_{1\nu}}/{k_1^2}\right),
\end{eqnarray}
where $\Gamma_z$ stands for total decay width of $Z^0$ boson. The parameter $e$ is the unit of electric charge and $g_w$ is the weak interaction coupling constant. $\varepsilon_{\sigma}(q_2)$ is the polarization vector of the final photon. $I^{\nu \sigma}$ is an one-loop tensor integral, which for the channel via the virtual photon, $e^+e^- \to \gamma^{*}/Z^{0} \to H(|c\bar{c}\rangle) +\gamma$, can be written as
\begin{widetext}
\begin{equation} \label{11}
I^{\nu \sigma}=\int \frac{f^{\nu \sigma}(k)~\textrm{d}^4k}{\left(k^2-m_c^2\right) \left[(k- q_1)^2-m_c^2\right] \left(k-\frac{q_1}{2}\right)^2 \left[(k-q_1-q_2)^2-m_c^2\right]}\;,
\end{equation}
\end{widetext}
where $k$ is the loop momentum and $f^{\nu \sigma}(k)$ is a tensor function of $k$. Production of different charmonium states $H(|c\bar{c}\rangle)$ shall result in different $f^{\nu \sigma}(k)$. For later usage of the region expansion method~\cite{shift}, we transfer Eq.(\ref{11}) into the following form by making a momentum shift for the loop momentum,
\begin{widetext}
\begin{equation} \label{12}
I^{\nu \sigma}=\int \frac{f^{\nu \sigma}(k+\frac{q_1}{2})~\textrm{d}^4k}{k^2\left[(k+\frac{q_1}{2})^2 -m_c^2\right] \left[(k- \frac{q_1}{2})^2-m_c^2\right]  \left[(k-\frac{q_1}{2}-q_2)^2-m_c^2\right]}\;,
\end{equation}
\end{widetext}
At the amplitude level, $f^{\nu \sigma}(k+\frac{q_1}{2})$ can be expanded in varies independent tensor structures, such as $k^{\nu} \varepsilon^{\sigma k q_1 q_2}$, the number of which depends on which charmonium states to be generated. The tensor structures have to be reduced to scalar forms before doing loop integration. All tensor integrals can be changed to scalar integrals through the reduction with the help of {\tt FeynCalc} package~\cite{FeynCalc}. For example, in $D$-dimension, the above mentioned tensor structure $k^{\nu} \varepsilon^{\sigma k q_1 q_2}$ can be reduced into the following form,
\begin{displaymath}
\frac{\varepsilon ^{\nu\sigma q_1 q_2} \left[k^2 m_c^2 (r-1)^2+k\cdot q_2 (k\cdot q_2-(r-1) k\cdot q_1)\right]}{m_c^2 (D-2) (r-1)^2}\;,
\end{displaymath}
where $r={s}/{(4 m_c^2)}$ and $s$ is the squared center-of-mass energy for $e^+ e^-$ collision. The results of {\tt FeynCalc} can be further simplified by using a Mathematica function {\tt \$Apart}~\cite{apart}, which is performed the simplification analytically at the amplitude level. By using {\tt \$Apart}, the integrals as Eq.(\ref{11}) can be decomposed into linear combinations of a number of standard integrals~\cite{feng}. For example, the box diagram Box N1 can be transformed into a linear superposition of four 2-point, five 3-point, and two 4-point scalar integrals corresponding to ${\cal N}^{(2)}_n$, ${\cal N}^{(3)}_n$ and ${\cal N}^{(4)}_n$, which are easier to be evaluated, i.e.
\begin{widetext}
\begin{eqnarray}\label{12}
I^{\nu \sigma}&=& \frac{2 \sqrt{2} i e^2 g_s^2\epsilon^{\nu \sigma q_1 q_2}}{9 m_c^2(D - 2)(r - 1)^2} \left\{2 m_c^4({D^2} - 12D + 28){(r - 1)^2} \left({\cal N}^{(4)}_1 + {\cal N}^{(4)}_2\right) - 4 r (D - 6)~{\cal N}^{(2)}_4 \right. \nonumber \\
&& \left. -({D^2} - 10D + 24) \left[m_c^2(r - 1) \left({\cal N}^{(3)}_1 + {\cal N}^{(3)}_2\right) + {\cal N}^{(3)}_4+{\cal N}^{(3)}_5\right]+ 2 (D - 6)(2 r - 1)~{\cal N}^{(2)}_3 \right. \nonumber \\
&& \left. + {(D - 6)^2} (r - 1)\left(2 m_c^2 r {\cal N}^{(3)}_3 + {\cal N}^{(2)}_2\right) + (6 - D)(D(r - 1) - 6 r + 4)~{\cal N}^{(2)}_1 \right\},
\end{eqnarray}
\end{widetext}
where $g_s^2=4 \pi \alpha_s$ is the strong coupling constant. The 2-point scalar integrals ${\cal N}^{(2)}_n$ (n=1,$\cdots$,4) are defined as following:
\begin{eqnarray}
{\cal N}^{(2)}_1&=& \lambda \int\frac{\textrm{d}^Dk}{k^2(k^2 + k \cdot q_1)} \;, \\
{\cal N}^{(2)}_2&=& \lambda \int\frac{\textrm{d}^Dk}{(k^2 + k \cdot q_1)(k^2 - k \cdot q_1)} \;, \\
{\cal N}^{(2)}_3&=& \lambda \int\frac{\textrm{d}^Dk}{k^2[(k-\frac{q_1}{2}-q_2)^2-m_c^2]} \;, \\
{\cal N}^{(2)}_4&=& \lambda \int\frac{\textrm{d}^Dk}{(k^2 + k \cdot q_1)[(k-\frac{q_1}{2}-q_2)^2-m_c^2]} \;,
\end{eqnarray}
where $\lambda= \frac{\mu^{2\epsilon}\Gamma(\epsilon)}{i(4\pi)^{D/2}}$. The 3-point scalar integrals ${\cal N}^{(3)}_n$ (n=1,$\cdots$,5) are defined as following:
\begin{eqnarray}
{\cal N}^{(3)}_1&=& \lambda \int\frac{\textrm{d}^Dk}{k^4(k^2 - k \cdot q_1)} \;, \\
{\cal N}^{(3)}_2&=& \lambda \int\frac{\textrm{d}^Dk}{k^4(k^2 + k \cdot q_1)} \;, \\
{\cal N}^{(3)}_3&=& \lambda \int\frac{\textrm{d}^Dk}{k^2(k^2 + k \cdot q_1)[(k-\frac{q_1}{2}-q_2)^2-m_c^2]} \;, \\
{\cal N}^{(3)}_4&=& \lambda \int\frac{k \cdot q_2~\textrm{d}^Dk}{k^4(k^2 - k \cdot q_1)} \;, \\
{\cal N}^{(3)}_5&=& \lambda \int\frac{k \cdot q_2~\textrm{d}^Dk}{k^4(k^2 + k \cdot q_1)} \;.
\end{eqnarray}
The 4-point scalar integrals ${\cal N}^{(4)}_1$ and ${\cal N}^{(4)}_2$ are formulated as
\begin{eqnarray}
{\cal N}^{(4)}_1&=& \lambda \int\frac{\textrm{d}^Dk}{{{k^4}({k^2} - k \cdot {q_1})[(k-\frac{q_1}{2}-q_2)^2-m_c^2]}} \;, \\
{\cal N}^{(4)}_2&=& \lambda \int\frac{\textrm{d}^Dk}{{{k^4}({k^2} + k \cdot {q_1})[(k-\frac{q_1}{2}-q_2)^2-m_c^2]}} \;.
\end{eqnarray}
To calculate those $n$-point scalar integrals ${\cal N}^{(n)}_m$, we reduce them into much simpler master integrals by using Feynman integral reduction algorithm {\tt FIRE}~\cite{fire}. The most important point of {\tt FIRE} is the so-called integration by parts (IBP) relations~\cite{ibp},
\begin{equation}
\int\cdots\int \textrm{d}^Dk_1 \textrm{d}^D k_2\cdots\frac{\partial}{\partial k_i}\left( p_j \frac{1}{E^{a_1}_1\cdots E^{a_n}_n}\right)=0\;,
\end{equation}
where $k_i$ is the loop momentum and $p_j$ can be either internal or external momentum in the concerned loop diagrams. Then, $I^{\nu \sigma}$ for the box diagram Box N1 can be reduced in terms of a linear combination of two 1-point, two 2-point and one 3-point master integrals, i.e.,
\begin{widetext}
\begin{eqnarray}
I^{\nu \sigma} &=& \frac{ i e^2 g_s^2 \epsilon^{\nu \sigma q_1 q_2}}{9\sqrt{2} m_c^4(D - 5)(D - 3)(s - 1)^2}\left(4m_c^2(D^2 - 8D + 15)\left[2r(m_c^2(D^2 - 13D + 38)(r - 1){\cal I}^{(3)}_{1} \right.\right. \nonumber \\
&& \left.\left. + ({D^2} - 13D + 36){\cal I}^{(2)}_{2}) - (D^2 - 13D + 36)(2r - 1){\cal I}^{(2)}_{1}\right] - (D - 2) \left(D^3(r + 1) \right.\right. \nonumber \\
&& \left.\left. - 18{D^2}(r + 1) + D(103r + 99) - 6(31r + 29)\right] {\cal I}^{(1)}_{1} + ({D^4} - 20{D^3} + 135{D^2} - 372D + 348)(r - 1){\cal I}^{(1)}_{2}\right),
\end{eqnarray}
\end{widetext}
where
\begin{eqnarray}
{\cal I}^{(1)}_1 &=& \lambda \int\frac{\textrm{d}^Dk}{k^2+ k \cdot q_1} \;,\\
{\cal I}^{(1)}_2 &=& \lambda \int\frac{\textrm{d}^Dk}{k^2- k \cdot q_1} \;,\\
{\cal I}^{(2)}_1 &=& \lambda \int\frac{\textrm{d}^Dk}{k^2[(k-\frac{q_1}{2}-q_2)^2-m_c^2]} \;,\\
{\cal I}^{(2)}_2 &=& \lambda \int\frac{\textrm{d}^Dk}{(k^2 + k \cdot q_1)[(k-\frac{q_1}{2}-q_2)^2-m_c^2]} \;, \\
{\cal I}^{(3)}_1 &=& \lambda \int\frac{\textrm{d}^Dk}{k^2(k^2 + k \cdot q_1)[(k-\frac{q_1}{2}-q_2)^2-m_c^2]} \;.
\end{eqnarray}
These master integrals can be easily evaluated as, for instance,
\begin{eqnarray}
{\cal I}^{(3)}_{1} &=& \lambda\int \frac{d^Dk}{k^2 \left[(k + \frac{q_1}{2})^2-m_c^2\right]\left[\left(k-\frac{q_1}{2}-q_2\right)^2-m_c^2\right]} \nonumber\\
&=& C_0\left[\left(\frac{q_1}{2}\right)^2,\left(q_1+q_2\right)^2, \left(\frac{q_1}{2}+ q_2\right)^2,0,m_c^2,m_c^2\right] \;, \nonumber
\end{eqnarray}
which can be numerically evaluated by using the package~{\tt LOOPTOOLS}~\cite{looptools}. In fact, analytic expression for the 3-point $C_0$ function can be obtained by using the method introduced by Refs.~\cite{thooft1,loop2}. And, we obtain
\begin{eqnarray}
&&C_0\left[\left(\frac{q_1}{2}\right)^2,\left(q_1+q_2\right)^2,\left(\frac{q_1}{2} +q_2\right)^2,0,m_c^2,m_c^2\right] \;,\nonumber\\
&=& C_0 \left[m_c^2, s, \frac{s}{2}-m_c^2, 0, m_c^2, m_c^2 \right] \;, \nonumber\\
&=& \frac{\lambda_1}{s-4 m_c^2}\left[2\textrm{Li}_2 \left(\frac{1}{2r-1} \right)+\ln^2(2r-1) -\frac{\pi ^2}{3}\right] \;,
\end{eqnarray}
where $\lambda_1 = \frac{i}{(4 \pi)^2}$ and $\text{Li}_2$ is the Spence function. Using the same techniques, one can evaluate all one-loop scalar integrals analytically, which are put in the Appendix for convenience. Finally, we obtain
\begin{widetext}
\begin{eqnarray}
I^{\nu \sigma} &=& \frac{e^2~g_s^2~\epsilon^{\nu \sigma q_1 q_2} |R_S(0)|} {72\sqrt{2}~\pi^{5/2}~m_c^{5/2}(r - 1)^2} \Bigg\{2(1 - r)\left[\ln \left(\frac{4\pi \mu^2}{m_c^2}\right)-2- \gamma_E\right] - 2r\left[\textrm{Li}_2(b) + \textrm{Li}_2(a) - \textrm{Li}_2\left(\frac{1}{2r - 1}\right)\right]  \nonumber\\
&& - r\left[\ln^2(b) + \ln^2(a) - \ln^2(2r - 1)\right] \Bigg\} - \frac{e^2~g_s^2~\epsilon^{\nu \sigma q_1 q_2} |R_S(0)|} {36\sqrt{2}~\pi^{5/2}~m_c^{5/2}(r - 1)} \frac{1}{\epsilon} \;,
\end{eqnarray}
\end{widetext}
where $a = \frac{1}{2}\left(1+\sqrt{\frac{r-1}{r}}\right)$ and $b = \frac{1}{2}\left(1-\sqrt{\frac{r-1}{r}}\right)$.

The amplitude of the channel $e^+ + e^-\to Z^0 \to H(|c\bar{c}\rangle)+\gamma$ can be treated via the same way as that of $e^+ + e^-\to \gamma^* \to H(|c\bar{c}\rangle)+\gamma$. As a subtle point for $e^+ + e^-\to Z^0 \to H(|c\bar{c}\rangle)+\gamma$, we have to deal with the $\gamma_5$ problem in using the dimensional regularization approach. We adopt the $\gamma^5$-scheme suggested in Ref.\cite{gamma5} to do our calculation, which has recently been applied by Refs.\cite{51,52,53,54,Wang20081}. Following the same treatments as described in detail in the appendix of Ref.\cite{Wang20081}, one needs to be careful about the following points:
\begin{itemize}
\item The cyclicity of the traces involving odd number of $\gamma^5$ should be treated by using the same route in order to keep the final finite results consistent, i.e. the route for summing up dummy index go across even or odd numbers of $\gamma_5$.
\item The amplitudes must be written starting from the same reading point, which ensures every amplitude dealing with upon the same gauge.
\item In order to guarantee the conservation of the vector current, the reading point must be the axial vector vertex for the case when the amplitude contains an anomalous axial current. While, for the cases of the amplitudes containing even number of $\gamma^5$, they are free from the $\gamma^5$-scheme, such as the productions of $\eta_c$ and $h_c$ via the process $e^++e^-\to Z^0 \to H(|c\bar{c}\rangle) +\gamma$.
\end{itemize}

Total cross section of the process can be written as
\begin{eqnarray}
\sigma = \langle{\cal O}^H(n) \rangle_0 \cdot \hat{\sigma}^{(0)}\left( 1+ \frac{\pi\alpha_s C_F}{v} + \frac{\alpha_s \hat{C}}{\pi} + {\cal O}(\alpha_s^2) \right),
\end{eqnarray}
where $\langle{\cal O}^H(n) \rangle_0$ is the tree-level non-perturbative but universal matrix element which represents the hadronization probability of the perturbative state $(c\bar{c})[n]$ into the bound state $H$. $\hat{\sigma}^{(0)}$ is the hard part of the LO cross sections. The Coulomb term ${\pi\alpha_s C_F}/{v}$ can be absorbed into the renormalized matrix element,
\begin{eqnarray}
\sigma = \langle{\cal O}^H(n) \rangle_{R} \cdot \hat{\sigma}^{(0)} \left( 1 + \frac{\alpha_s \hat{C}}{\pi} + {\cal O}(\alpha_s^2) \right) ,
\end{eqnarray}
where $\langle{\cal O}^H(n) \rangle_{R}$ is the redefined matrix elements at the one-loop level. The color-singlet matrix elements can be related with the wavefunction at the origin for $S$-wave charmonium or the first derivative of the wavefunction at the origin for $P$-wave charmonium~\cite{nrqcd,NRQCD2},
\begin{displaymath}
\frac{\langle 0|{\cal O}_{\bf 1}^{\eta_c}|0 \rangle_{R}}{2N_c} \simeq \frac{\langle 0|{\cal O}_{\bf 1}^{J/\psi}|0 \rangle_{R}}{6N_c} = |\Psi_{1S}(0)|^2
\end{displaymath}
and
\begin{displaymath}
\frac{\langle 0|{\cal O}_{\bf 1}^{\chi_{c0}}|0 \rangle_{R}}{2N_c} \simeq \frac{\langle 0|{\cal O}_{\bf 1}^{\chi_{c1}}|0 \rangle_{R}}{6N_c} \simeq \frac{\langle 0|{\cal O}_{\bf 1}^{\chi_{c2}}|0 \rangle_{R}}{10N_c} =|\Psi'_{1P}(0)|^2 .
\end{displaymath}
There values can be derived by using the potential model, e.g., Ref.~\cite{pot}, or be extracted from the corresponding charmonium decays by comparing with the data, c.f. Refs.~\cite{Wang20082,prdDL,prd84034022}.

\section{Numerical results}\label{sec3}

\subsection{Input parameters}

If not specially stated, the input parameters are taken as the same as those of Ref.\cite{zcharmlo}. We take $m_z=91.1876$ GeV and the charm quark mass $m_c=1.5$ GeV, $\alpha(10.6\;{\rm GeV})=1/130.9$, and the charmonium masses~\cite{pdg}: $M_{J/\psi}=3.097$, $M_{\eta_{c}}=2.980$, $M_{\chi_{c0}}=3.415$, $M_{\chi_{c1}}=3.511$, and $M_{\chi_{c2}}=3.556$ GeV, respectively. We adopt two-loop strong coupling constant to do our calculation, i.e.,
\begin{equation}
\alpha_s(\mu_R) = \frac{4\pi}{\beta_0 L} - \frac{4\pi \beta_1 \ln(L)}{\beta_0^3 L^2} \;,
\end{equation}
where $L=\ln(\mu_R^2/\Lambda_{\textrm{QCD}}^2)$, $\beta_0=11-\frac{2}{3} n_f$, and $\beta_1=\frac{2}{3}(153-19 n_f)$. The active flavor number $n_f=4$ and $\Lambda^{(4)}_{\textrm{QCD}}=0.332$ GeV. The central renormalization scale $\mu_R$ is chosen to be $2m_c$.

The color-singlet non-perturbative matrix elements are related to the wavefunction at the origin $|\Psi_S(0)|={|R_S(0)|}/{\sqrt{4\pi}}$ for $S$-wave state and its first derivative at the origin  $|\Psi'_P(0)| =\sqrt{\frac{3}{4\pi}} |R'_P(0)|$ for $P$-wave state. Based on the experimental values for the leptonic width of $J/\psi$ and the width of $\chi_{c2}$ to two photons, we can inversely determine the radial wavefunction $|R_{ns}(0)|$ at the origin and the first derivative of the radial wavefunction at the origin $|R^{'}_{np}(0)|$ at the LO or NLO level through the formulas up to NLO level~\cite{pdg}
\begin{equation}
\Gamma_{\psi(ns)\rightarrow e^+e^-} = {4\alpha^2\over 9m^2_c}\left(1-{16\alpha_s(2m_c) \over 3\pi}\right)|R_{ns}(0)|^2 \label{wave1}
\end{equation}
and
\begin{equation}
\Gamma_{\chi(np)\rightarrow \gamma\gamma} = {64\alpha^{2}\over 45m^{4}_{c}}\left(1-{16\alpha_{s}(2m_c) \over 3\pi}\right)|R^{'}_{np}(0)|^{2}. \label{wave2}
\end{equation}
For experimental values of these decay widths, we adopt those from the Particle Data Group~\cite{pdg}: $\Gamma_{J/\psi\rightarrow e^+e^-} =(5.55\pm0.16)$ keV and $\Gamma_{\chi_{c_2}\rightarrow \gamma\gamma}=(0.514\pm0.062)$ keV. Then, as a combined error of both the squared average of the experimental errors on the decay widths and the theoretical errors caused by varying the scale within the conventional region of $[m_c,4m_c]$, we obtain~\cite{pmc1}
\begin{eqnarray}
|R_{J/\psi}(0)|^{2}_{\rm LO} &=& \left(0.481^{+0.014}_{-0.013}\right) \;{\rm GeV}^{3} ,\label{pot1} \\
|R^{'}_{\chi_{cJ}}(0)|^{2}_{\rm LO} &=& \left(0.031^{+0.004}_{-0.004}\right) \;{\rm GeV}^{5} \label{pot2}
\end{eqnarray}
at the LO level; and
\begin{eqnarray}
|R_{J/\psi}(0)|^{2}_{\rm NLO} &=& \left(0.855^{+0.044}_{-0.051}\right) \;{\rm GeV}^{3} , \label{potnlo1} \\
|R^{'}_{\chi_{cJ}}(0)|^{2}_{\rm NLO} &=& \left(0.056^{+0.007}_{-0.007}\right) \;{\rm GeV}^{5} \label{potnlo2}
\end{eqnarray}
at the NLO level.

As a cross check of our present calculation up to one-loop level, we find that: (I) Both UV and IR divergences are canceled exactly when summing all divergent terms together, which is checked both analytically and numerically. (II) When taking the same input parameters, we obtain the same LO estimations for all the mentioned channels as those of Ref.\cite{zcharmlo} which are calculated by using the improved trace technology. (III) As for the experimental condition of $B$ factory, when taking the same input parameters as those of Ref.\cite{higher-charm}, we obtain the same numerical results for the channel $e^+ e^- \to \gamma^* \to H(|c\bar{c}\rangle)+\gamma$ up to NLO level.

\subsection{Basic results}

\begin{table}
\begin{tabular}{|c|c|c|c|c|}
\hline
  & &\multicolumn{3}{|c|}{NLO}\\
  \cline{3-5}
\raisebox{1.6ex}[0pt]{$\sqrt{s}=10.6~\textrm{GeV}$}& ~~\raisebox{1.6ex}[0pt]{LO}~~ & ~~Born~~ & ~~Virt.~~ & ~~$R_X$~~ \\
\hline
$\sigma_{\gamma^* \to \eta_{c} \gamma} $ & 40.3 & 71.7 & $-15.1$ & $-21\%$ \\
\hline
$\sigma_{\gamma^* \to \chi_{c0} \gamma} $ & 0.77 & 1.38 & 0.25 & 18\%\\
\hline
$\sigma_{\gamma^* \to \chi_{c1} \gamma} $ & 8.55 & 15.4 & $-4.23$ & $-27\%$\\
\hline
$\sigma_{\gamma^* \to \chi_{c2} \gamma} $ & 3.36 & 6.08 & $-4.48$ & $-74\%$\\
\hline
\end{tabular}
\caption{Total cross sections (in unit: fb) for the charmonium production at LO and NLO levels in $\alpha_s$ at the $B$ factories with $\sqrt{s}=10.6$ GeV. $\mu_R=2m_c$. Born stands for the NLO Born terms and Virt. stands for the virtual correction. } \label{crosectY}
\end{table}

\begin{table}
\begin{tabular}{|c|c|c|c|c|}
\hline
 & &\multicolumn{3}{|c|}{NLO}\\
 \cline{3-5}
\raisebox{1.6ex}[0pt]{~~$\sqrt{s}=m_Z$~~} & ~~\raisebox{1.6ex}[0pt]{LO}~~ & ~~Born~~ & ~~Virt.~~ & ~~$R_X$~~ \\
\hline
$\sigma_{Z^0 \to \eta_{c} \gamma} $ & 0.431 & 0.774 & 0.146 & 19\% \\
\hline
$\sigma_{Z^0 \to \emph{J}/\psi \gamma}$ & 2.987 & 5.309 & 1.378 & 26\% \\
\hline
$\sigma_{Z^0 \to h_{c} \gamma} $ & 0.257 & 0.463 & $-0.307$ & $-66\%$ \\
\hline
$\sigma_{Z^0 \to \chi_{c0} \gamma}$ & 0.013 & 0.023 & $2.50\times10^{-4}$ & 1.1\% \\
\hline
$\sigma_{Z^0 \to \chi_{c1} \gamma}$ & 0.076 & 0.135 & 0.020  & 15\%  \\
\hline
$\sigma_{Z^0 \to \chi_{c2} \gamma}$ & 0.025 & 0.045 & $-0.035$& $-78\%$  \\
\hline
\end{tabular}
\caption{Total cross sections (in unit: fb) for the charmonium production at LO and NLO levels in $\alpha_s$ at the super $Z$ factory with $\sqrt{s}=m_z$ GeV. $\mu_R=2m_c$. Born stands for the NLO Born terms and Virt. stands for the virtual correction. }\label{crosectZ}
\end{table}

We present the LO and NLO results for the channels $e^+e^-\to \gamma^*/Z^0 \to H+ \gamma$ with the center-of-mass collision energies $\sqrt{s}=10.6$ GeV and $\sqrt{s}=m_Z$ in Table \ref{crosectY} and \ref{crosectZ}, respectively. At the $B$ factory, we only consider the channel via a virtual photon, since its cross section dominates the channel via $Z^0$ boson by at least four orders. For the similar reason, at the super $Z$ factory, we only consider the channel via $Z^0$ boson \footnote{In addition, it has been shown that the interference terms between the channel via a virtual photon and the channel via $Z^0$ boson only lead to small contributions at both the $B$ and super $Z$ factories~\cite{zcharmlo}. So, in the present paper, we will not consider the interference terms either. }.

It is noted that our present LO cross sections are smaller than those listed in Ref.\cite{zcharmlo}. It is because that at present, we adopt $|R(0)|_{\rm LO}$ or $|R'(0)|_{\rm LO}$ to be those of Eqs.(\ref{pot1},\ref{pot2}) other than the value derived by potential model~\cite{pot} to do our LO calculation. To be consistent, the NLO cross sections are calculated by using the NLO values $|R(0)|_{\rm NLO}$ and $|R'(0)|_{\rm NLO}$ that are presented in Eqs.(\ref{potnlo1},\ref{potnlo2}). Further more, we define a ratio $R_X$ to show the relative importance of the Born terms and the virtual corrections at the NLO level,
\begin{equation}
R_X = \left.\frac{\sigma_{\rm Virt.}}{\sigma_{\rm Born}}\right\vert_{X} ,
\end{equation}
where $X$ stands for specific production channel. Since the real correction contributes zero, $R_{X}$ rightly shows the pQCD convergence up to NLO level. The magnitude of $|R_{X}|$ for all production channels are sizable, which shows the necessity and importance of one-loop corrections. The virtual contributions are either positive or negative depending on which charmonium state to be generated and which channel to be adopted. At the $B$ factory, we see that in most cases the one-loop QCD corrections are negative and moderate, except for the $\chi_{c2}$ case, in which the correction is large and is about $-74\%$ of the Born result, consistent with those of Ref.\cite{higher-charm}.  At the super $Z$ factory, we see that in most cases the one-loop QCD corrections are positive and moderate, except for the $h_c$, $\chi_{c0}$ and $\chi_{c2}$ cases, in which the corrections are large and are about $-66\%$, $1.1\%$ and $-76\%$ of the Born results, respectively.

\begin{figure}[tb]
\includegraphics[width=0.5\textwidth]{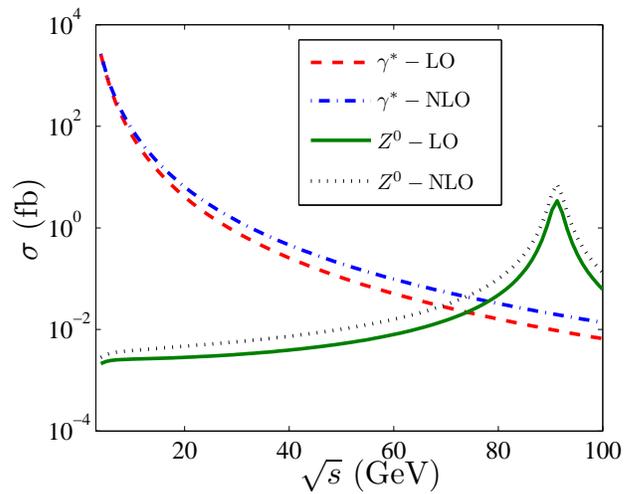}
\caption{Total cross sections versus the $e^{+}e^{-}$ collision energy $\sqrt{s}$ for the channels via the virtual photon and intermediate $Z^0$ boson $e^+e^-\to \gamma^*/\ Z^0 \to H(|c\bar{c}\rangle)+\gamma$. The label $\gamma^*-{\rm LO}$ means the channel via a virtual photon and at the LO level, and etc.. }\label{ecm1}
\end{figure}

To give an idea of how the total cross sections change with the $e^+e^-$ collision energy, we present both the LO and the NLO results in Fig.~\ref{ecm1}. For convenience, in Fig.~\ref{ecm1}, all color-singlet charmonium states' contributions are summed up for the same production channel. The summation of all Fock states are reasonable, since higher Fock state may decay to the ground state via strong or electromagnetic radiations with high probability. Up to NLO level, because of $Z^0$-boson resonance effect, there is a peak value at $\sqrt{s}=m_{Z}$. Thus, one may expect that those channels via $Z^0$ boson can provide sizable contributions at the super $Z$ factory.
\begin{table}
\begin{tabular}{|c||c|c|c|}
\hline
~~ $\mu_R$ ~~ & ~~ $2m_c$ ~~ & ~~$\frac{\sqrt{s}}{2}$~~ & ~~$\sqrt{s}$~~ \\
\hline
\hline
$\sigma_{\gamma^* \to \eta_{c} \gamma} $ & 56.6 & 59.4 & 61.6 \\
\hline
$\sigma_{\gamma^* \to \chi_{c0} \gamma} $ & 1.63 & 1.59 & 1.55\\
\hline
$\sigma_{\gamma^* \to \chi_{c1} \gamma} $ & 11.2 & 12.0 & 12.6\\
\hline
$\sigma_{\gamma^* \to \chi_{c2} \gamma} $ & 1.60 & 2.42 & 3.07\\
\hline
\hline
$\sigma_{Z^0 \to \eta_{c} \gamma} $ & 0.929 & 0.853 & 0.845 \\
\hline
$\sigma_{Z^0 \to \emph{J}/\psi \gamma} $ & 6.687 & 5.989 & 5.914 \\
\hline
$\sigma_{Z^0 \to h_{c} \gamma} $ & 0.157 & 0.312 & 0.329 \\
\hline
$\sigma_{Z^0 \to \chi_{c0} \gamma} $ & 0.023 & 0.023 & 0.023\\
\hline
$\sigma_{Z^0 \to \chi_{c1} \gamma} $ & 0.157 & 0.147 & 0.145\\
\hline
$\sigma_{Z^0 \to \chi_{c2} \gamma} $ & 0.010 & 0.028 & 0.030\\
\hline
\end{tabular}
\caption{Scale uncertainties of total cross sections (in unit: fb) at NLO level caused by three different choices of $\mu_R$. Here, for the $e^+e^-$ annihilation channel via a virtual photon, $\sqrt{s}=10.6$ GeV; for the channel via $Z^0$, $\sqrt{s}=m_Z$. $m_c=1.5$ GeV. } \label{mu}
\end{table}

\begin{figure}[htb]
\includegraphics[width=0.45\textwidth]{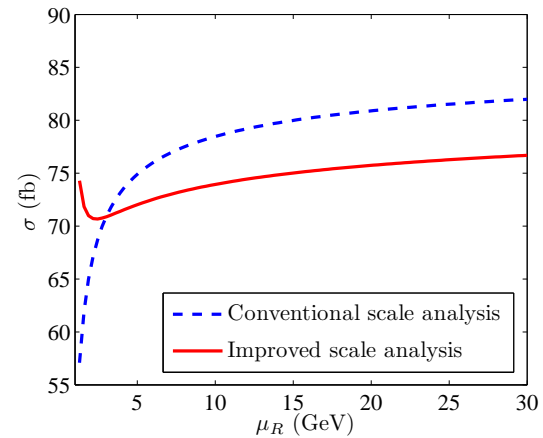}
\includegraphics[width=0.45\textwidth]{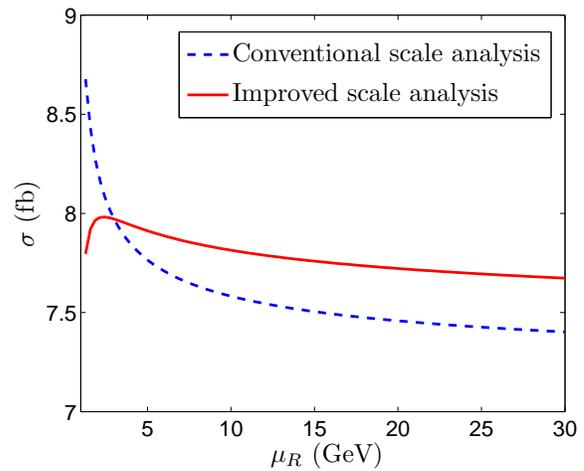}
\caption{Total cross sections versus the renormalization scale $\mu_R$ at the NLO level for $e^+e^-\to \gamma^* \to H(|c\bar{c}\rangle)+\gamma$ (upper) and $e^+e^-\to Z^0 \to H(|c\bar{c}\rangle)+\gamma$ (lower), in which all color-singlet charmonium states have been summed up. Two methods for scale analysis are adopted.} \label{mu1}
\end{figure}

As a subtle point, how to set optimal renormalization scale for the process is an important issue in making precise pQCD predictions. Conventionally, one may choose one typical energy scale $Q$ and then vary it within the region of $[Q/2,2Q]$ or directly choose several typical energy scales for probing the scale dependence. As an example, we present the total cross sections under three frequently used scales, i.e., $\mu_R=2m_c$, $\frac{\sqrt{s}}{2}$, and $\sqrt{s}$, in Table \ref{mu}. More generally, we present total cross sections versus the scale $\mu_R$ up to $30$ GeV in Fig.~\ref{mu1}, in which all color-singlet charmonium states have been summed up. Under the conventional scale setting, by varying $\mu_R\in[2m_c,\sqrt{s}]$, a large scale uncertainty about $11\%$ (or $-8.6\%$) is observed for the channel via a virtual (or via a $Z^0$ boson).

In addition to the conventional way of doing scale uncertainty, in Fig.~\ref{mu1}, we also present the results for an improved way to estimate the scale uncertainty, which is suggested in Ref.\cite{pmc1} and is based on the principle of maximum conformality (PMC)~\cite{pmc} \footnote{The PMC is programmed to eliminate the scale uncertainty at any fixed order by absorbing all non-conformal terms $\beta$-terms into the running coupling. Since those $\beta$-terms rightly determine the running behavior via renormalization group equation, one can obtain the optimal behavior of the running coupling and hence obtain the optimal scale of the process.}. According to the suggestion, even though we have no $\beta$-terms to determine the optimal scale, we can compensate the conventional scale uncertainty at the NLO level by using one-higher order terms from the $\alpha_s$ running known from the renormalization group equation, and then a more reliable scale analysis can be achieved. That is, we substitute the following formulae
\begin{equation}\label{eqas2}
\alpha_s(2m_c) = \alpha_s(\mu_R) \left[1 - \alpha_s(\mu_R)\frac{\beta_0}{4 \pi} \ln\left(\frac{4 m_c^2}{\mu_R^2} \right)\right]
\end{equation}
into the NLO expressions of $e^+e^-\to \gamma^* \to H(|c\bar{c}\rangle)+\gamma$. Fig. \ref{mu1} really shows a better scale uncertainty than that of the conventional one, i.e. by varying $\mu_R\in[2m_c,\sqrt{s}]$, a smaller scale uncertainty about $4.5\%$ (or $-5.2\%$) is observed for the channel via a virtual (or via a $Z^0$ boson).

\begin{figure*}
\includegraphics[width=0.45\textwidth]{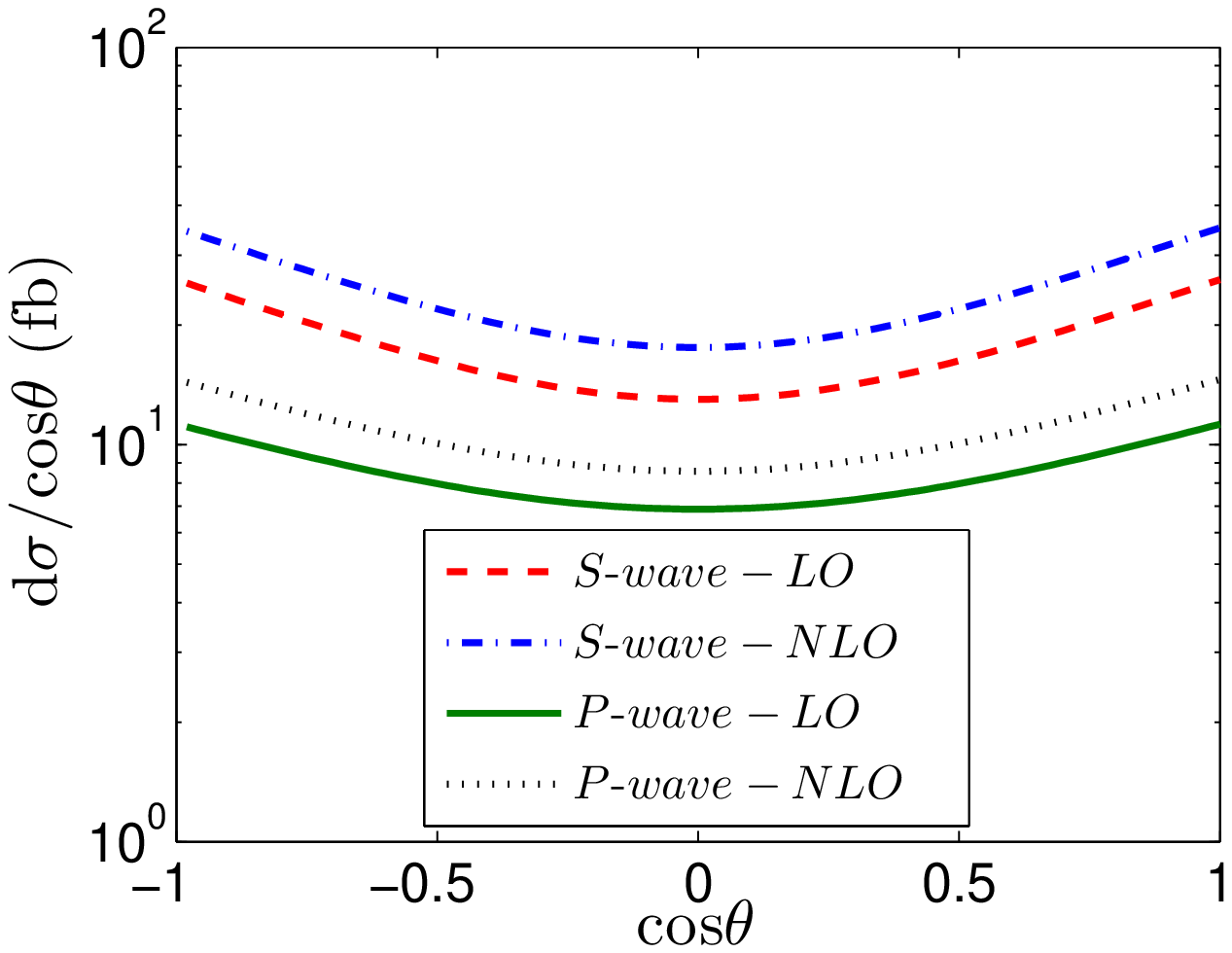}
\includegraphics[width=0.45\textwidth]{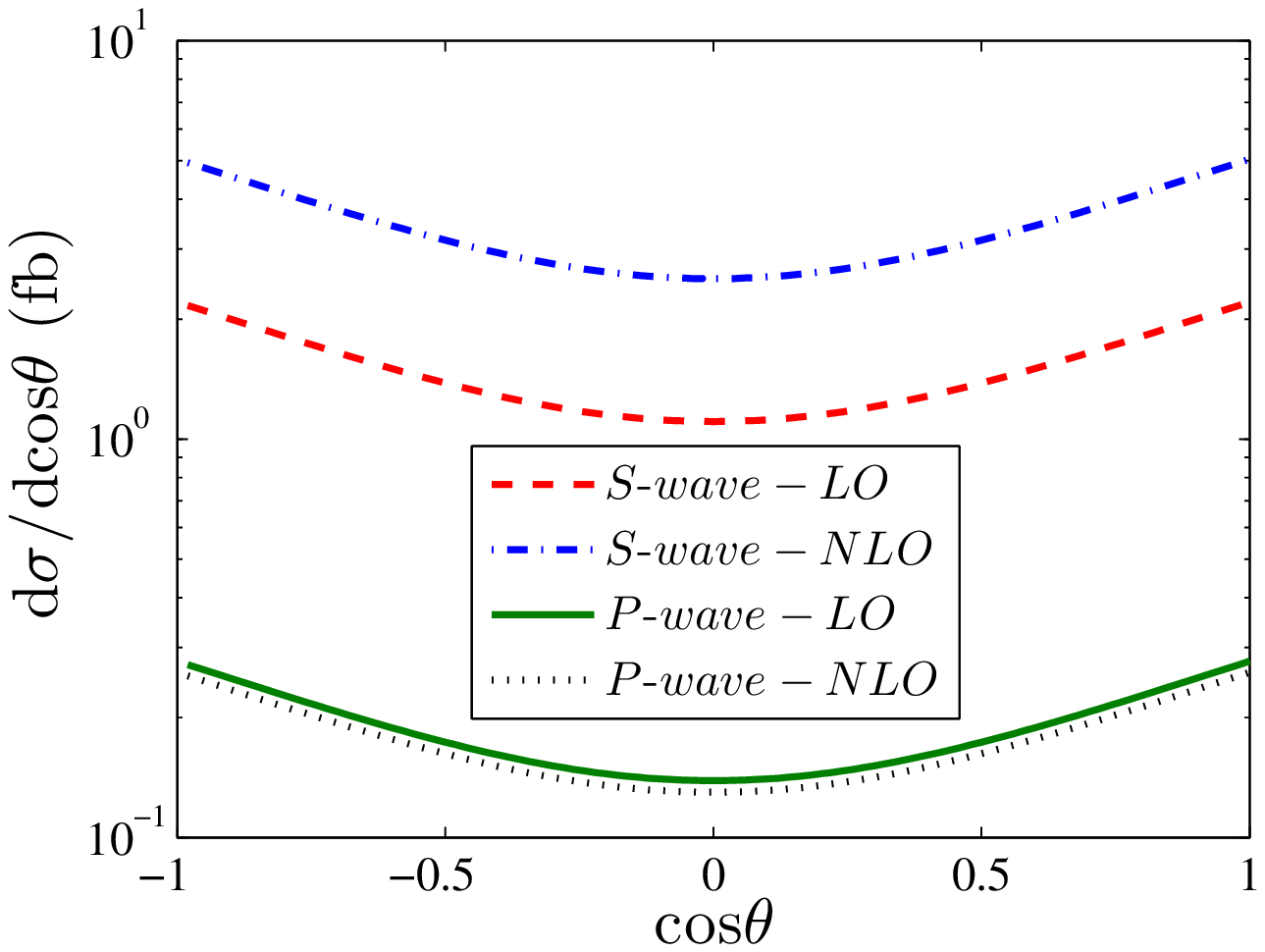}
\caption{Differential cross sections $d\sigma/d\cos\theta$ for $e^+e^-\to \gamma^*/Z^0 \to H(|c\bar{c}\rangle)+\gamma$ up to NLO level with $\sqrt{s}=10.6$ GeV (left) and $\sqrt{s}=m_Z$ (right), respectively. The contributions from all color-singlet charmonium states have been summed up separately for $S$ and $P$ wave states accordingly. } \label{diff1}
\end{figure*}

\begin{figure*}
\includegraphics[width=0.45\textwidth]{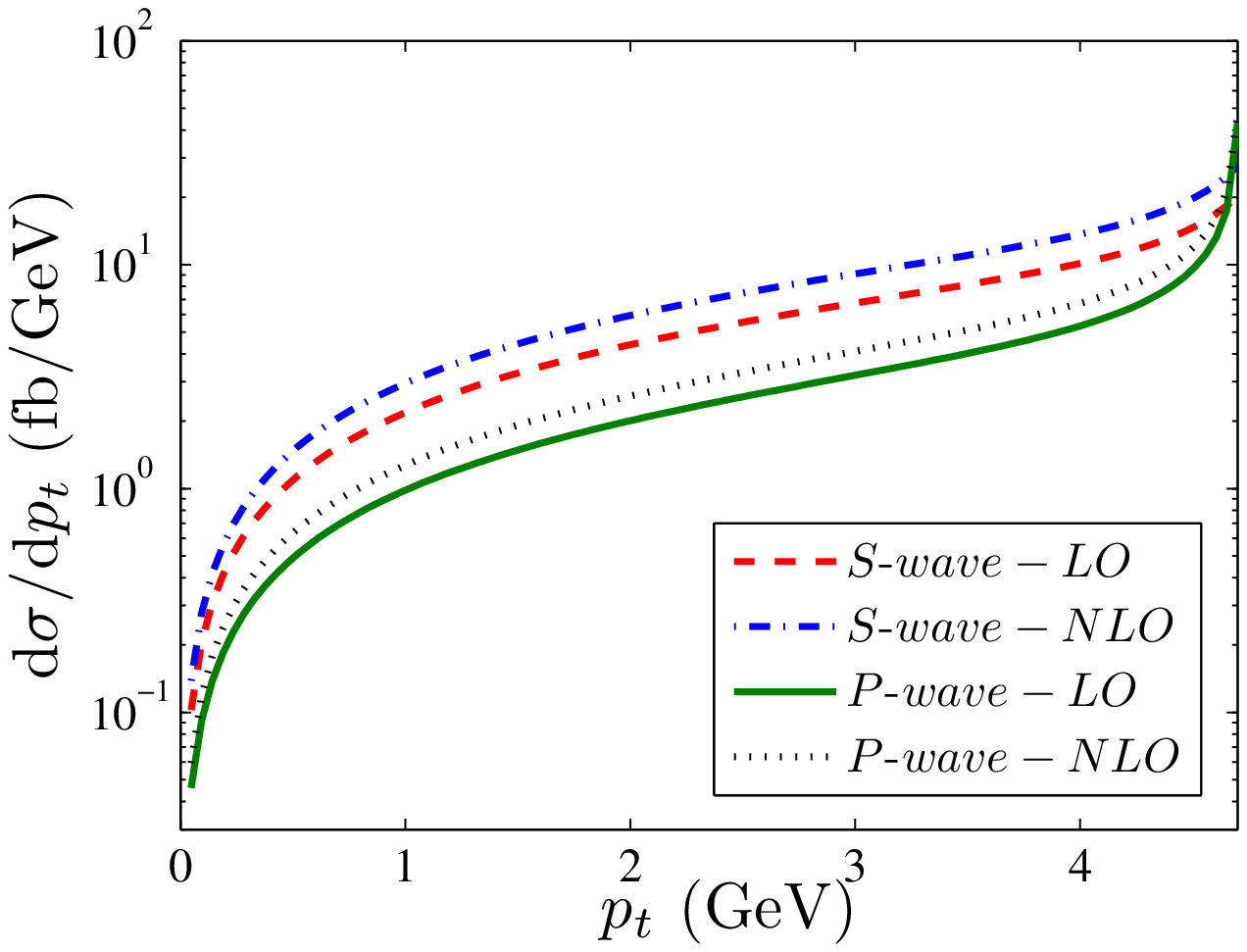}
\includegraphics[width=0.45\textwidth]{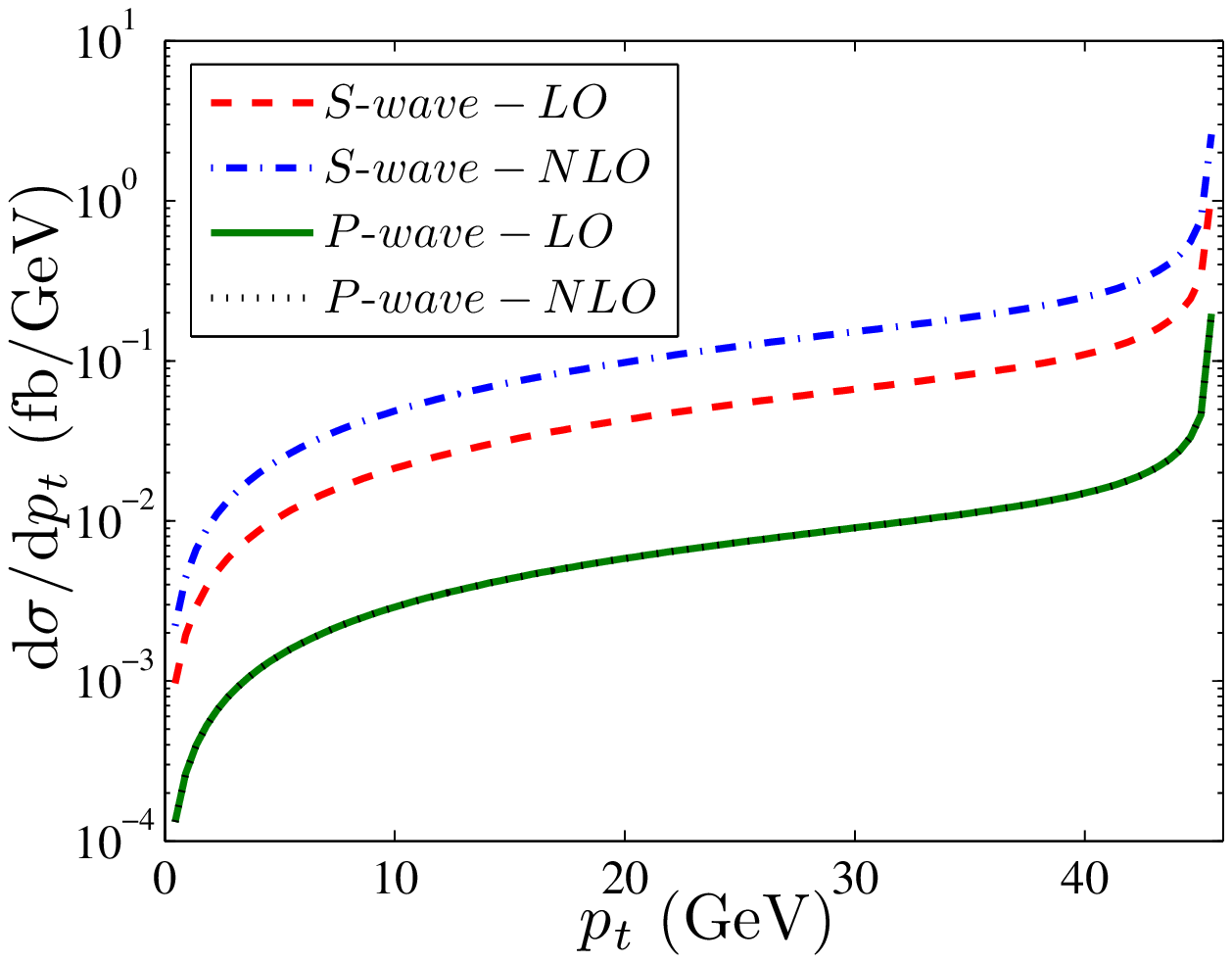}
\caption{Differential cross sections $d\sigma/dp_t$ for $e^+e^-\to \gamma^*/Z^0 \to H(|c\bar{c}\rangle)+\gamma$ up to NLO level with $\sqrt{s}=10.6$ GeV (left) and $\sqrt{s}=m_Z$ (right), respectively. The contributions from all color-singlet charmonium states have been summed up separately for $S$ and $P$ wave states accordingly. } \label{diff2}
\end{figure*}

We put the differential distributions $d\sigma/d\cos\theta$ and $d\sigma/dp_t$ up to NLO level in Figs.~\ref{diff1} and \ref{diff2}, where $\theta$ stands for the angle between the three-vector momentums of the charmonium and initial electron and $p_t$ is the charmonium transverse momentum. Here to show how the one-loop correction affect the production, we put the $S$-wave and $P$-wave cross sections in a separate way. Because the phase spaces are the same for both the LO and the NLO terms, the NLO distributions have the same shapes as those of LO distributions, but their differences are sizable. Then, a NLO calculation is necessary to achieve a more accurate estimation.

\begin{table}
\begin{tabular}{|c||c|c|c|c|}
\hline
 $\sqrt{s}\;(\textrm{GeV}$) & 10.6 & 91.1876 & 125 & 500 \\
\hline
\hline
$\sigma_{\gamma^* \to \eta_{c} \gamma} $ & 59.4 & 0.017 & 0.005 & $2.3\times10^{-5}$ \\
\hline
$\sigma_{\gamma^* \to \chi_{c0} \gamma} $ & 1.63 & $4.2\times10^{-4}$ & $1.1\times10^{-4}$ & $4.3\times10^{-7}$\\
\hline
$\sigma_{\gamma^* \to \chi_{c1} \gamma} $ & 11.2 & 0.003 & $8.2\times10^{-4}$ & $4.0\times10^{-6}$\\
\hline
$\sigma_{\gamma^* \to \chi_{c2} \gamma} $ & 1.60 & $1.9\times10^{-4}$ & $4.9\times10^{-5}$ & $1.5\times10^{-7}$\\
\hline
\hline
$\sigma_{Z^0 \to \eta_{c} \gamma} $ & $4.4\times10^{-4}$ & 0.929 & $8.8\times10^{-4}$ & $1.0\times10^{-6}$\\
\hline
$\sigma_{Z^0 \to J/\psi \gamma} $ & 0.004 & 6.687 & 0.006 & $7.3\times10^{-6}$\\
\hline
$\sigma_{Z^0 \to h_{c} \gamma} $ & $1.4\times10^{-4}$ & 0.157 & $1.4\times10^{-4}$ & $1.1\times10^{-7}$\\
\hline
$\sigma_{Z^0 \to \chi_{c0} \gamma} $ & $1.3\times10^{-5}$ & 0.023 & $2.0\times10^{-5}$ & $1.9\times10^{-8}$\\
\hline
$\sigma_{Z^0 \to \chi_{c1} \gamma} $ & $8.7\times10^{-5}$ & 0.157 & $1.5\times10^{-4}$ & $1.8\times10^{-7}$\\
\hline
$\sigma_{Z^0 \to \chi_{c2} \gamma} $ & $1.2\times10^{-5}$ & 0.010 & $8.8\times10^{-6}$ & $6.4\times10^{-9}$\\
\hline
\end{tabular}
\caption{Total cross sections (in unit: fb) at NLO level for the process $e^+e^-\to \gamma^*/Z^0 \to H(|c\bar{c}\rangle)+ \gamma$ with different center-of-mass collision energy, here $\mu_R=2m_c$ and $m_c=1.5~\textrm{GeV}$ are adopted. $\sqrt{s}=10.6$, 91.1876, 125, and 500 GeV correspond to $B$ factories, the super $Z$ factory, the Higgs factory, and the ILC, respectively.}
\label{events}
\end{table}

As a final remark, we make a comparison of single charmonium production at different experimental conditions suggested in the literature, the results of which are listed in Table \ref{events}. It indicates that
\begin{itemize}
\item At the $B$ factories with an integrated luminosity up to 20fb$^{-1}$, when summing all the charmonium states'contributions together, about $1.5\times10^3$ total charmonium events could be observed.
\item Supposing the high luminosity ${\cal L}\simeq 10^{36}$cm$^{-2}$s$^{-1}$ at the super $Z$ factory, when summing all charmonium states' contribution together, there are total about $8.0\times10^4$ charmonium events to be generated in one operation year. More specifically, we shall have $9.3\times10^3$ $\eta_c$, $6.7\times10^4$ $J/\psi$, $1.6\times10^3$ $h_c$, $230$ $\chi_{c0}$, $1.6\times10^3$ $\chi_{c1}$, and $100$ $\chi_{c2}$, respectively. Then, such super $Z$ factory will provide an ideal platform to make precise measurements charmonium properties, even for some of its excited states, via its dominant production channel $e^+e^-\to Z^0 \to H(|c\bar{c}\rangle)+\gamma$.
\item Due to the very small cross sections in the Higgs factory ($\sqrt{s}=125$ GeV) and ILC ($\sqrt{s}=500$ GeV or higher)~\cite{ILC}, there are almost no events to be obtained even under a high integrated luminosity up to 100fb$^{-1}$. Thus, in different to the $B$ factory and the super $Z$ factory, the suggested Higgs factory and the ILC are not suitable for observing the charmonium events via the production channel $e^+e^- \to H(|c\bar{c}\rangle)+\gamma$.
\end{itemize}

\section{Summary}\label{sec4}

The measured double charmonium production cross sections are unexpectedly large in comparison with the LO calculation~\cite{BLB2003,BL2003,plB557}. Even though many suggestions have been tried to explain the puzzle~\cite{prd67054007,prd77014002,sum rule,Braguta2008,plb57039,NLO corrections, relativistic, BLC2008, Wang20081,Wang20082,wang2012,wang2013}, it has not been well settled so far. It is therefore helpful to find another channel such as the single charmonium production or another experimental platform other than the $B$ factory to check all theoretical estimations and treatments. It is noted that the charmonium production at BESIII may also be helpful for such purpose~\cite{chao}.

It has been shown that the super $Z$ factory running with high luminosity ${\cal L}\simeq 10^{36}$cm$^{-2}$s$^{-1}$ can provide a potential platform to study the single charmonium production via $e^+ e^- \to \gamma^*/Z^0 \to H(c\bar{c})+\gamma$~\cite{zcharmlo}. In the present paper, we have presented an improved analysis up to NLO level, i.e. both the LO and NLO estimations at the $B$ factories and the super $Z$ factory are discussed. The NLO distributions have the same shapes as those of LO distributions, but their differences are sizable, which inversely indicates that a NLO calculation is necessary and important to achieve an accurate estimation.

At the $B$ factory, we see that in most cases the one-loop QCD corrections are negative and moderate. While for the super $Z$ factory, the one-loop QCD corrections are positive and moderate in most cases. At the super $Z$ factory with a high luminosity up to ${\cal L}\simeq 10^{36}$cm$^{-2}$s$^{-1}$, one may observe about $8.0\times10^{4}$ charmonium events via the channel $e^+e^-\to Z^0 \to H(|c\bar{c}\rangle)+\gamma$ in one operation year. Renormalization scale uncertainties have been discussed with an improved treatment based on the idea of PMC scale setting. We have shown that by using such improved treatment, part of the conventional scale uncertainty is compensated by the one higher-order running behavior of the strong coupling constant, then a smaller scale uncertainty than the conventional scale analysis have been observed. That is, by varying $\mu_R\in[2m_c,\sqrt{s}]$, about a smaller $4.5\%$ (or $-5.2\%$) is observed for the channel via a virtual (or via a $Z^0$ boson).

\hspace{2cm}

{\bf Acknowledgement:} We thank Wen-Long Sang for helpful discussions. This work was supported in part by the Fundamental Research Funds for the Central Universities under Grant No.CQDXWL-2012-Z002, by Natural Science Foundation of China under Grant No.11075225 and No.11275280.

\appendix

\begin{widetext}

\section{Analytical results for one-loop integrals}

We present the analytical expressions for the six independent scalar integrals needed in our present calculations. There is only one independent 1-point scalar integral,
\begin{eqnarray}
{\cal I}^{(1)}&=& \lambda \int\frac{\textrm{d}^Dk}{\left(k\pm\frac{q_1}{2} \right)^2-m_c^2}=\lambda \int\frac{\textrm{d}^Dk}
{\left(k\pm\frac{q_1}{2}\pm q_2\right)^2-m_c^2} = A_0(m_c^2) \;.
\end{eqnarray}
There are two independent 2-point scalar integrals,
\begin{eqnarray}
{\cal I}^{(2)}_{1}&=& \lambda \int\frac{\textrm{d}^Dk}{k^2 \left[\left(k\pm\frac{q_1}{2}\pm q_2\right)^2-m_c^2 \right]} = B_0\left[\left(\frac{q_1}{2}+q_2\right)^2,0,m_c^2\right] \;,\\
{\cal I}^{(2)}_{2}&=& \lambda \int\frac{\textrm{d}^Dk}{\left(k^2-m_c^2\right) \left[\left(k-q_1- q_2\right)^2-m_c^2 \right]} = B_0\left[\left(q_1 + q_2 \right)^2,m_c^2,m_c^2\right] \;.
\end{eqnarray}
And there are three independent 3-point scalar integrals,
\begin{eqnarray}
{\cal I}^{(3)}_{1} &=& \lambda \int\frac{\textrm{d}^Dk}{k^2 \left[\left(k+\frac{q_1}{2}\right)^2-m_c^2 \right]\left[\left(k - \frac{q_1}{2} - q_2\right)^2-m_c^2 \right]} = C_0\left[\left(\frac{q_1}{2}\right)^2,\left(-q_1-q_2\right)^2,\left(-\frac{q_1}{2}-q_2\right)^2,0,m_c^2,m_c^2\right] \;,\\
{\cal I}^{(3)}_{2} &=& \lambda \int\frac{\textrm{d}^Dk}{k^2 \left[\left(k - \frac{q_1}{2}\right)^2-m_c^2 \right]\left[\left(k - \frac{q_1}{2} - q_2\right)^2-m_c^2 \right]} = C_0\left[\left(-\frac{q_1}{2}\right)^2,\left(-q_2\right)^2,\left(-\frac{q_1}{2}-q_2\right)^2,0,m_c^2,m_c^2\right] \;,\\
{\cal I}^{(3)}_{3} &=& \lambda \int\frac{\textrm{d}^Dk}{\left(k ^2-m_c^2 \right) \left[\left(k - q_1\right)^2-m_c^2 \right] \left[\left(k - q_1 - q_2\right)^2-m_c^2 \right]} = C_0\left[\left(-q_1\right)^2,\left(-q_2\right)^2,\left(-q_1-q_2\right)^2,m_c^2,m_c^2,m_c^2\right] \;.
\end{eqnarray}
Here $A_0$, $B_0$ and $C_0$ are conventional 1-point, 2-point, and 3-point scalar functions, respectively. More specifically, after simplification, we have
\begin{eqnarray}
{\cal I}^{(1)} &=&\lambda_1 m_c^2 \left[\ln\left(\frac{\mu^2}{m_c^2} \right)+\frac{1}{\epsilon }-\gamma_{E} +1 \right]\;,\\
{\cal I}^{(2)}_{1} &=& \lambda_1 \left[\ln \left(\frac{\mu^2}{m_c^2}\right)-\frac{2r-2 }{2 r-1}\ln (2r-2)+\frac{1}{\epsilon }-\gamma_{E} +2 \right] \;,\\
{\cal I}^{(2)}_{2} &=& \lambda_1 \left[ \ln \left(\frac{\mu^2}{s}\right)+(a-b)\ln \left(\frac{b}{a}\right)-\ln(a b)+ \frac{1}{\epsilon }-\gamma_E +2 \right] \;,\\
{\cal I}^{(3)}_{1} &=& \frac{\lambda_1}{s-4 m_c^2}\left[2\textrm{Li}_2\left(\frac{1}{2r-1}\right)+\ln ^2(2r-1)-\frac{\pi ^2}{3}\right]\;,\\
{\cal I}^{(3)}_{2} &=& \frac{\lambda_1}{s-4 m_c^2}\left[2 \textrm{Li}_2 (a) +2 \textrm{Li}_2(b)-2 \textrm{Li}_2\left(\frac{1}{2 r-1}\right)+\ln^2(a)+\ln^2(b)-\ln ^2(2r-1)\right]\;,\\
{\cal I}^{(3)}_{3} &=& \frac{\lambda_1}{s-4 m_c^2}\left[\textrm{Li}_2(a)+ \textrm{Li}_2(b) + \frac{\ln^2(a)}{2} + \frac{\ln^2(b)}{2} - \frac{\pi ^2}{6} \right]\;,
\end{eqnarray}
where
\begin{eqnarray}
\lambda_1 = \frac{i}{(4 \pi)^2},\;
r = \frac{s}{4m_c^2},\;
a = \frac{1}{2}\left(1+\sqrt{\frac{r-1}{r}}\right),\;
b = \frac{1}{2}\left(1-\sqrt{\frac{r-1}{r}}\right).\;
\end{eqnarray}

\end{widetext}

\end{document}